\documentclass[12pt]{amsart}
\usepackage{amssymb, amsthm, amsfonts, latexsym}
\usepackage[leqno]{amsmath} 
\usepackage{natbib}
\usepackage{mathtools}
\usepackage{enumitem}
\usepackage{needspace}
\usepackage{array}
\usepackage{graphicx,booktabs,tabularx,longtable,pdflscape,float}
\usepackage{xurl}
\usepackage[multiple]{footmisc}
\usepackage{comment}

\newcommand{\CM}{C^{v}_{\mcircled{M}}}
\newcommand{\CG}{C^{v}_{G}}
\newcommand{\CR}{C^{v}_{R}}
\newcommand{\Cv}{C^{v}}
\newcommand{\nv}{n^{v}}
\newcommand{\gdom}{\succeq^{G}}
\newcommand{\M}{\mathcal{M}}

\usepackage[left=1.25in, top=1.25in, right=1.25in, bottom=1.25in]{geometry}
\usepackage{setspace}
\usepackage{tgpagella}
\usepackage{microtype}

\usepackage{xcolor}
\usepackage{tikz}

\newcommand{\mcircled}[1]{%
  \mathord{\tikz[baseline=(char.base)]{\node[draw,circle,inner sep=0.5pt] (char) {$\scriptstyle #1$};}}%
}

\usepackage{etoolbox}
\patchcmd{\section}{\scshape}{\bfseries}{}{}
\makeatletter
\renewcommand{\@secnumfont}{\bfseries}
\makeatother

\theoremstyle{plain}

\newtheorem{lemma}{Lemma}

\newtheorem*{proposition*}{Proposition}

\theoremstyle{definition}

\newtheoremstyle{boldremark}
 {0.5\topsep}{0.5\topsep}
 {\normalfont}{}
 {\bfseries}{.}
 {.5em}{}
\theoremstyle{boldremark}

\usepackage[
 colorlinks=true,
 linkcolor=blue,
 citecolor=blue,
 urlcolor=blue,
 filecolor=blue,
 hyperfootnotes=false,
 hyperindex,
 breaklinks
]{hyperref}

\begin{document}
\raggedbottom

\author[Aygün]{Orhan Aygün$^\textrm{1}$}\thanks{$^\textrm{1}$ Department of Economics, Boğaziçi University. Email: \href{orhan.aygun@bogazici.edu.tr}{orhan.aygun@bogazici.edu.tr}}

\author[Turhan]{Bertan Turhan$^\textrm{2}$}\thanks{$^\textrm{2}$ Department of Economics, Iowa State University. Email: \href{bertan@iastate.edu}{bertan@iastate.edu}}

\date{\today}

\title{Merit-Optimality and Elongation of Transversal Matroids}
\date{\today}

\begin{abstract}
\onehalfspacing
We introduce the (s)-fold elongation of a transversal matroid as a new analytical tool for matching theory. By using this tool, we provide an alternative proof to establish that the meritorious horizontal choice rule of \cite{sonmez2022affirmative} is merit-optimal. 
\end{abstract}

\begin{titlepage}
    \maketitle
\end{titlepage}

\section{Introduction}

Matroid theory has become an important analytical framework in matching theory and market design, particularly in allocation problems that combine merit-based selection with distributional objectives. While the existing literature has extensively exploited the structural properties of matroids, certain operations on matroids remain largely unexplored in economic design. In this paper, we introduce the $s$-fold elongation of transversal matroids (\cite{higgs1968strong} and \cite{Oxley2011}) as a potentially powerful analytical tool for matching theory and market design.

The central idea is to embed a selection problem with distributional objectives into an enlarged matroid that accommodates additional selections beyond those necessary to maximize the underlying distributional protections, while preserving the combinatorial structure that permits greedy optimization. This approach allows us to study seemingly distinct stages of a choice procedure within a unified matroid-theoretic framework.

To demonstrate the usefulness of this methodology, we revisit the merit-optimality of the meritorious horizontal choice rule established by \cite{sonmez2022affirmative}. We show that their two-step choice rule can be represented as the greedy choice rule of an appropriately elongated transversal matroid and that every selection that maximally accommodates horizontal reservation protections is independent in this matroid. Merit-optimality then follows directly from the lemma of \cite{gale1968optimal}, without requiring an inductive argument.

The methodological contribution of this paper extends beyond this particular application. Matroid elongation offers a systematic way to reconcile distributional objectives with merit-based selection while retaining the structural properties necessary for greedy optimization. We anticipate that this methodology will prove useful in a broader class of matching and allocation problems involving affirmative action, diversity objectives, reserve systems, and other distributional considerations.

\section{Framework}\label{sec:setup}

We follow the notation of \citet{sonmez2022affirmative}, henceforth SY. Fix a vertical category $v \in \mathcal{V}$ with capacity $q^{v}$, a set of traits $\mathcal{T}$ with HR quotas $(q^{v}_{t})_{t \in \mathcal{T}}$ satisfying $\sum_{t \in \mathcal{T}} q^{v}_{t} \le q^{v}$, and the set $\mathcal{I}^{v}$ of individuals eligible for category-$v$ positions, with distinct merit scores $\sigma$.

For $I \subseteq \mathcal{I}^{v}$, the category-$v$ HR graph and the HR-maximality function $\nv(I)$ are as in SY. $\nv$ is the rank function of the \emph{transversal matroid} with ground set $\mathcal{I}^{v}$ induced by the HR graph \citep{edmonds1965transversals}. We use the standard properties of matroid rank functions as in R1--R3 of SY Appendix A.1, together with the unit-increase property $\nv(X \cup \{e\}) \le \nv(X) + 1$, which follows from R1 and R3.

A \emph{single-category choice rule} is a map $\Cv\colon 2^{\mathcal{I}^{v}} \to 2^{\mathcal{I}^{v}}$ with $\Cv(I) \subseteq I$ and $|\Cv(I)| \le q^{v}$ for every $I$. It \emph{maximally accommodates HR protections} if, for every $I \subseteq \mathcal{I}^{v}$ and every $j \in I \setminus \Cv(I)$, $\nv\bigl(\Cv(I) \cup \{j\}\bigr) = \nv\bigl(\Cv(I)\bigr)$.

The \emph{meritorious horizontal choice rule} $\CM$ is defined in SY Section 4.2.3. Its Step~1 is the greedy choice rule of the transversal matroid with merit scores as weights. Individuals are considered in decreasing order of merit score, and an individual is selected if and only if adding her increases the rank of the set selected so far. We write $\CG(I)$ for its output, so that $\CG(I)$ is independent and $|\CG(I)| = \nv(\CG(I)) = \nv(I)$. Step~2 then adds the $\min\{q^{v},|I|\} - \nv(I)$ highest merit-score members of $I \setminus \CG(I)$. We write $\CR(I)$ for this set, so that
\[
\CM(I) = \CG(I) \cup \CR(I) \qquad\text{and}\qquad |\CM(I)| = \min\{q^{v}, |I|\}.
\]
By SY Lemma 2, $\CG$ satisfies: if $i \in \CG(I)$ and $i \in I' \subseteq I$, then $i \in \CG(I')$.\footnote{SY Lemma 2 is shown in \cite{fleiner2001matroid}.}

Finally, recall Gale domination \citep{gale1968optimal}: for $X, Y \subseteq \mathcal{I}^{v}$, write $X \gdom Y$ if $|X| \ge |Y|$ and, for every $k \le |Y|$, the $k$th highest merit-score individual of $X$ has a weakly higher merit score than the $k$th highest merit-score individual of $Y$.\footnote{Since merit scores are distinct, position-wise weak inequality means that either the two individuals coincide, or the one in $X$ has the strictly higher score.}

\begin{proposition*}[\textbf{Proposition 2 of SY}]
    Given a category $v \in \mathcal{V}$, let $\Cv$ be any single-category choice rule that maximally accommodates HR protections. Then, for every $I \subseteq \mathcal{I}^{v}$,
\begin{itemize}
\item[(1)] $|\Cv(I)| \le |\CM(I)|$, and
\item[(2)] for every $k \le |\Cv(I)|$, if $i$ is the $k$th highest merit-score individual in $\CM(I)$ and $j$ is the $k$th highest merit-score individual in $\Cv(I)$, then $i = j$ or $\sigma(i) > \sigma(j)$.
\end{itemize}
\end{proposition*}

\section{Elongation of Transversal Matroids}\label{sec:fix}

The horizontal-reservation choice problem can be embedded into an appropriately enlarged matroid. We consider the $s$-fold elongation of the transversal matroid induced by the HR graph, where $s$ is the number of positions filled in the second step of the meritorious horizontal choice rule. We show that every selection that maximally accommodates HR protections is independent in this enlarged matroid, whereas the meritorious horizontal choice is its greedy base under the merit ordering. Thus, the merit-domination property follows directly from Gale's lemma, without requiring induction or a separate analysis of the individuals selected in the two steps. Our approach shows how matroid elongation can accommodate additional selections beyond those needed to maximize HR protections while preserving the structure required for greedy optimization. The same construction may be useful in other problems that combine merit-based selection with distributional constraints.

We show that, at every instance, $\CM(I)$ is the greedy base of a suitably enlarged matroid in which every maximally accommodating selection is independent. Proposition 2 then follows from a single application of Gale's lemma.

Fix $I \subseteq \mathcal{I}^{v}$ throughout this section, and let
\[
s = \min\{q^{v}, |I|\} - \nv(I) \;\ge\; 0,
\]
where nonnegativity uses $\nv(I) \le \sum_{t} q^{v}_{t} \le q^{v}$ and $\nv(I) \le |I|$. Define
\begin{equation}\label{eq:elong}
\M \;=\; \Bigl\{\, X \subseteq I \;:\; |X| \le \nv(X) + s \,\Bigr\}.
\end{equation}
In the language of matroid theory, $(I, \M)$ is the $s$-fold elongation of the transversal matroid restricted to $I$ (\citealp{higgs1968strong} and see also Section 7.3 of \citealp{Oxley2011}). For completeness, we verify directly that it is a matroid.

We first record a standard property of rank functions.

\begin{lemma}\label{lem:sat}
Let $r$ be a matroid rank function and $A \subseteq B$. If $r(A \cup \{e\}) = r(A)$ for every $e \in B \setminus A$, then $r(B) = r(A)$.
\end{lemma}

\begin{proof}
We show the claim by induction on $|B \setminus A|$. The case $|B\setminus A| \le 1$ is trivial. Suppose that the claim holds up to $m$, and let $|B \setminus A| = m+1$. Let $B = B' \cup \{e\}$ with $e \in B \setminus A$ and $|B' \setminus A| = m$, so $r(B') = r(A)$ by the induction hypothesis. Submodularity (R3) applied to $B'$ and $A \cup \{e\}$ gives
\[
r(B) + r(A) \;\le\; r(B') + r(A \cup \{e\}) \;=\; 2\,r(A),
\]
while monotonicity (R2) gives $r(B) \ge r(A)$. Hence $r(B) = r(A)$.
\end{proof}

\begin{lemma}\label{lem:matroid}
$(I, \M)$ is a matroid. Its bases are exactly the sets $X \in \M$ with $|X| = \min\{q^{v}, |I|\}$, and $\CM(I)$ is one such base.
\end{lemma}

\begin{proof}
\emph{(M1)} $\emptyset \in \M$ since $s \ge 0$.

\emph{(M2)} Let $X \in \M$ and $X' \subseteq X$, and suppose toward a contradiction that $|X'| > \nv(X') + s$. By the unit-increase property, $\nv(X) \le \nv(X') + |X \setminus X'|$. Then
\[
|X| \;=\; |X'| + |X \setminus X'| \;>\; \nv(X') + s + |X \setminus X'| \;\ge\; \nv(X) + s,
\]
contradicting $X \in \M$.

\emph{(M3)} Let $X_{1}, X_{2} \in \M$ with $|X_{1}| < |X_{2}|$. If $|X_{1}| < \nv(X_{1}) + s$, then for any $e \in X_{2} \setminus X_{1}$ (nonempty since $|X_2|>|X_1|$), monotonicity gives $|X_{1} \cup \{e\}| = |X_1|+1 \le \nv(X_{1}) + s \le \nv(X_{1} \cup \{e\}) + s$, so $X_{1} \cup \{e\} \in \M$. If instead $|X_{1}| = \nv(X_{1}) + s$, then
\[
\nv(X_{2}) \;\ge\; |X_{2}| - s \;>\; |X_{1}| - s \;=\; \nv(X_{1}),
\]
so $\nv(X_{1} \cup X_{2}) \ge \nv(X_{2}) > \nv(X_{1})$, and the contrapositive of Lemma \ref{lem:sat} (with $A = X_{1}$, $B = X_{1} \cup X_{2}$) yields $e \in X_{2} \setminus X_{1}$ with $\nv(X_{1} \cup \{e\}) = \nv(X_{1}) + 1$. Then $|X_{1} \cup \{e\}| = \nv(X_{1} \cup \{e\}) + s$, so $X_{1} \cup \{e\} \in \M$.

\emph{Bases.} Every $X \in \M$ satisfies $|X| \le \nv(X) + s \le \nv(I) + s = \min\{q^{v}, |I|\}$, by monotonicity. Conversely, $|\CM(I)| = \min\{q^{v}, |I|\}$, and $\CG(I) \subseteq \CM(I) \subseteq I$ together with $\nv(\CG(I)) = \nv(I)$ and monotonicity gives $\nv\bigl(\CM(I)\bigr) = \nv(I)$, so
\[
|\CM(I)| \;=\; \nv(I) + s \;=\; \nv\bigl(\CM(I)\bigr) + s ,
\]
hence $\CM(I) \in \M$. Thus, the maximum cardinality of a member of $\M$ is $\min\{q^{v}, |I|\}$ and is attained. Since all bases of a matroid have the same cardinality, the bases of $(I,\M)$ are precisely the members of $\M$ of that cardinality, $\CM(I)$ among them.
\end{proof}

\begin{lemma}\label{lem:Cindep}
If the choice rule $\Cv$ maximally accommodates HR protections, then $\Cv(I) \in \M$.
\end{lemma}

\begin{proof}
Applying Lemma \ref{lem:sat} with $A = \Cv(I)$, $B = I$, and $r = \nv$ gives $\nv\bigl(\Cv(I)\bigr) = \nv(I)$. Since $\Cv$ is a choice rule, $$|\Cv(I)| \le \min\{q^{v}, |I|\} = \nv(I) + s = \nv\bigl(\Cv(I)\bigr) + s.$$
\end{proof}

The next lemma is the pointwise content of SY Lemma 3. 

\begin{lemma}\label{lem:pointwise}
Let $(E, \mathcal{N})$ be a matroid with distinct weights $w$, and let $B^{*}$ be the base selected by the greedy algorithm on $E$. Let $B$ be a base such that, for every $e_{1} \in B$ and $e_{2} \in E \setminus B$ with $w(e_{2}) > w(e_{1})$, the set $(B \setminus \{e_{1}\}) \cup \{e_{2}\}$ is not a base. Then $B = B^{*}$.
\end{lemma}

\begin{proof}
Suppose $B \ne B^{*}$. Since bases are equicardinal, there exists $e_{1} \in B \setminus B^{*}$. By the symmetric single-element exchange property (the strengthening of property B2 given in SY Appendix A.1, due to \citealp{brualdi1969comments}), there exists $e_{2} \in B^{*} \setminus B$ such that both $(B \setminus \{e_{1}\}) \cup \{e_{2}\}$ and $(B^{*} \setminus \{e_{2}\}) \cup \{e_{1}\}$ are bases. By Gale's lemma (SY Lemma 1), $B^{*} \gdom (B^{*} \setminus \{e_{2}\}) \cup \{e_{1}\}$.

Summing the position-wise inequalities of Gale domination over the two equicardinal sets gives $\sum_{e \in B^{*}} w(e) \ge \sum_{e \in B^{*}} w(e) - w(e_{2}) + w(e_{1})$. That is, $w(e_{2}) \ge w(e_{1})$, and hence $w(e_{2}) > w(e_{1})$ since weights are distinct and $e_{1} \ne e_{2}$. But then $(B \setminus \{e_{1}\}) \cup \{e_{2}\}$ is a base with $e_{1} \in B$, $e_{2} \in E \setminus B$, and $w(e_{2}) > w(e_{1})$, which contradicts the hypothesis on $B$.
\end{proof}

\begin{lemma}\label{lem:greedy}
$\CM(I)$ is the base of $(I, \M)$ selected by the greedy algorithm with weights $\sigma$.
\end{lemma}

\begin{proof}
By Lemma \ref{lem:matroid}, $\CM(I)$ is a base of $(I, \M)$. By Lemma \ref{lem:pointwise}, it suffices to show the following: if $i \in \CM(I)$ and $j \in I \setminus \CM(I)$ with $\sigma(j) > \sigma(i)$, then $X = \bigl(\CM(I) \setminus \{i\}\bigr) \cup \{j\}$ is not a base of $(I, \M)$. Since $|X| = |\CM(I)| = \nv(I) + s$, by \eqref{eq:elong} and Lemma \ref{lem:matroid} the set $X$ is a base if and only if $\nv(X) \ge \nv(I)$. We show that $\nv(X) = \nv(I) - 1$.

First, $i \in \CG(I)$, and every $e \in \CR(I) \cup \{j\}$ satisfies $e \in I \setminus \CG(I)$ and $\sigma(e) > \sigma(i)$. To see this, note that $j \in I \setminus \CG(I)$ was not selected in Step~2, which selects the highest merit-score members of $I \setminus \CG(I)$. Hence every member of $\CR(I)$ has a higher merit score than $j$, and therefore than $i$. In particular $i \notin \CR(I)$, so $i \in \CG(I)$.

Second, let $A = \CG(I) \setminus \{i\}$, and fix $e \in \CR(I) \cup \{j\}$. Let $G_{e}$ be the set of members of $\CG(I)$ with a higher merit score than $e$. Since $e$ was not selected by the greedy rule of Step~1, $\nv(G_{e} \cup \{e\}) = \nv(G_{e})$. Since $\sigma(e) > \sigma(i)$, we have $i \notin G_{e}$, so $G_{e} \subseteq A$. Submodularity (R3) applied to $A$ and $G_{e} \cup \{e\}$, whose intersection is $G_{e}$ because $e \notin A$, gives
\[
\nv(A \cup \{e\}) + \nv(G_{e}) \;\le\; \nv(A) + \nv(G_{e} \cup \{e\}) \;=\; \nv(A) + \nv(G_{e}),
\]
so $\nv(A \cup \{e\}) = \nv(A)$ by monotonicity.

Finally, $X = A \cup \CR(I) \cup \{j\}$, so Lemma \ref{lem:sat} applied to $A \subseteq X$ gives $\nv(X) = \nv(A)$. As a subset of the independent set $\CG(I)$, the set $A$ is independent, so $\nv(A) = |A| = \nv(I) - 1$. Hence $X$ is not a base.
\end{proof}

\section*{Proof of Proposition}
Part (1) is immediate, exactly as in SY. For part (2), fix $I$ and consider the matroid $(I, \M)$ of \eqref{eq:elong}. By Lemma \ref{lem:greedy}, $\CM(I)$ is the greedy base of $(I, \M)$. By Lemma \ref{lem:Cindep}, $\Cv(I)$ is an independent set of $(I, \M)$. Gale's lemma (SY Lemma 1), applied in the matroid $(I, \M)$, yields $\CM(I) \gdom \Cv(I)$.

Position-wise, and using the distinctness of merit scores, this says that for every $k \le |\Cv(I)|$, the $k$th highest merit-score individual $i$ of $\CM(I)$ and the $k$th highest merit-score individual $j$ of $\Cv(I)$ satisfy $i = j$ or $\sigma(i) > \sigma(j)$.

Note that the matroid $(I, \M)$ depends on the instance $I$ through $s$. This is harmless, because Gale's lemma is applied at a fixed instance.

\bibliography{bertan}
\bibliographystyle{aer}

\end{document}